\documentclass[a4paper]{article}

\usepackage{INTERSPEECH2020}
\usepackage{subfigure}
\usepackage{multirow}
\usepackage{array}
\usepackage{color}
\newcolumntype{P}[1]{>{\centering\arraybackslash}p{#1}}

\title{iMetricGAN: Intelligibility Enhancement for Speech-in-Noise using Generative Adversarial Network-based Metric Learning}
\name{Haoyu Li$^1{^,}^2$, Szu-Wei Fu$^3$, Yu Tsao$^3$, Junichi Yamagishi$^1{^,}^2$}
\address{
  $^1$National Institute of Informatics, Japan  
  $^2$SOKENDAI, Japan\\
  $^3$Academia Sinica, Taiwan}
\email{haoyuli@nii.ac.jp, d04922007@ntu.edu.tw, yu.tsao@citi.sinica.edu.tw, jyamagis@nii.ac.jp}
\begin{document}

\maketitle
\begin{abstract}
The intelligibility of natural speech is seriously degraded when exposed to adverse noisy environments. 
In this work, we propose a deep learning-based speech modification method to compensate for the intelligibility loss, with the constraint that the root mean square (RMS) level and duration of the speech signal are maintained before and after modifications. Specifically, we utilize an \mbox{iMetricGAN} approach to optimize the speech intelligibility metrics with generative adversarial networks (GANs). Experimental results show that the proposed \mbox{iMetricGAN} outperforms conventional state-of-the-art algorithms in terms of objective measures, i.e., speech intelligibility in bits (SIIB) and extended short-time objective intelligibility (ESTOI), under a Cafeteria noise condition. In addition, formal listening tests reveal significant intelligibility gains when both noise and reverberation exist.

\end{abstract}

\section{Introduction}

Speech is the main media used by humans to communicate in daily life. However, in speech application systems, the intelligibility of speech messages is inevitably degraded due to background noise and reverberation. Several modification algorithms have been studied to enhance the intelligibility by preprocessing a signal before it is played out \cite{Chermaz2019}. This task is usually termed \textit{near-end listening enhancement} (NELE). 
There are various strategies for NELE tasks, such as spectral tilt flattening ~\cite{tilt-shift}, formant shifting ~\cite{nathwani2016formant}, and dynamic range compression ~\cite{drc1,drc2}. A common idea of the above algorithms is to reallocate the speech energy in the time-frequency (T-F) domain in such a way as to boost the acoustic cues that are perceptually crucial. 

In this paper, we utilize deep neural networks (DNNs) to reallocate the speech energy. The output of a DNN acts as scale factor $\alpha$, which is then multiplied to the T-F bin of the speech. The T-F bin energy is boosted with $\alpha > 1$ and suppressed otherwise. This framework is very similar to the masking-based speech enhancement approach \cite{narayanan2013ideal}, where spectral mask is predicted by NN and applied to the T-F bin as well. Although NELE task shares a similar solution to DNN-based speech enhancement, very few related works have been done so far. One of the biggest challenges with the DNN-based NELE approach is that there is no ground truth label that can be provided for supervised training. Specifically, given an unmodified plain speech, there is no standard that explicitly defines what the perfectly intelligible speech should be, and thus no ground truth label can be prepared. In contrast, in a speech enhancement task, clean speech without noise mixed can be easily prepared and regarded as the training label of corresponding noisy speech.

Recently, MetricGAN \cite{fu2019metricgan} was proposed and shown to be effective in optimizing the evaluation metrics in the field of speech enhancement. Inspired by its success, we adapt it to a modified \textit{\mbox{iMetricGAN}} that fits in the intelligibility enhancement task. \mbox{iMetricGAN} is a generative adversarial network (GAN) system that consists of a generator to enhance the speech signal as the intelligibility enhancement module and a discriminator to learn to predict the intelligibility scores of modified speech. Instead of discriminating fake from real, the discriminator aims to closely approximate the intelligibility metrics as a learned surrogate, and then the generator can be trained properly with the guidance of this surrogate. From another point of view, with the framework of \mbox{iMetricGAN}, the ground truth speech label can be implicitly defined as a modified speech that achieves the maximum value of the intelligibility scores predicted by the discriminator. Consequently, the proposed \mbox{iMetricGAN} can effectively optimize the intelligibility of speech even though no ground truth label is provided. Furthermore, \mbox{iMetricGAN} is a flexible language-independent framework that can be easily extended to optimize multiple metrics simultaneously.

\begin{figure}[tbp]
  \centering
  \includegraphics[width=\linewidth]{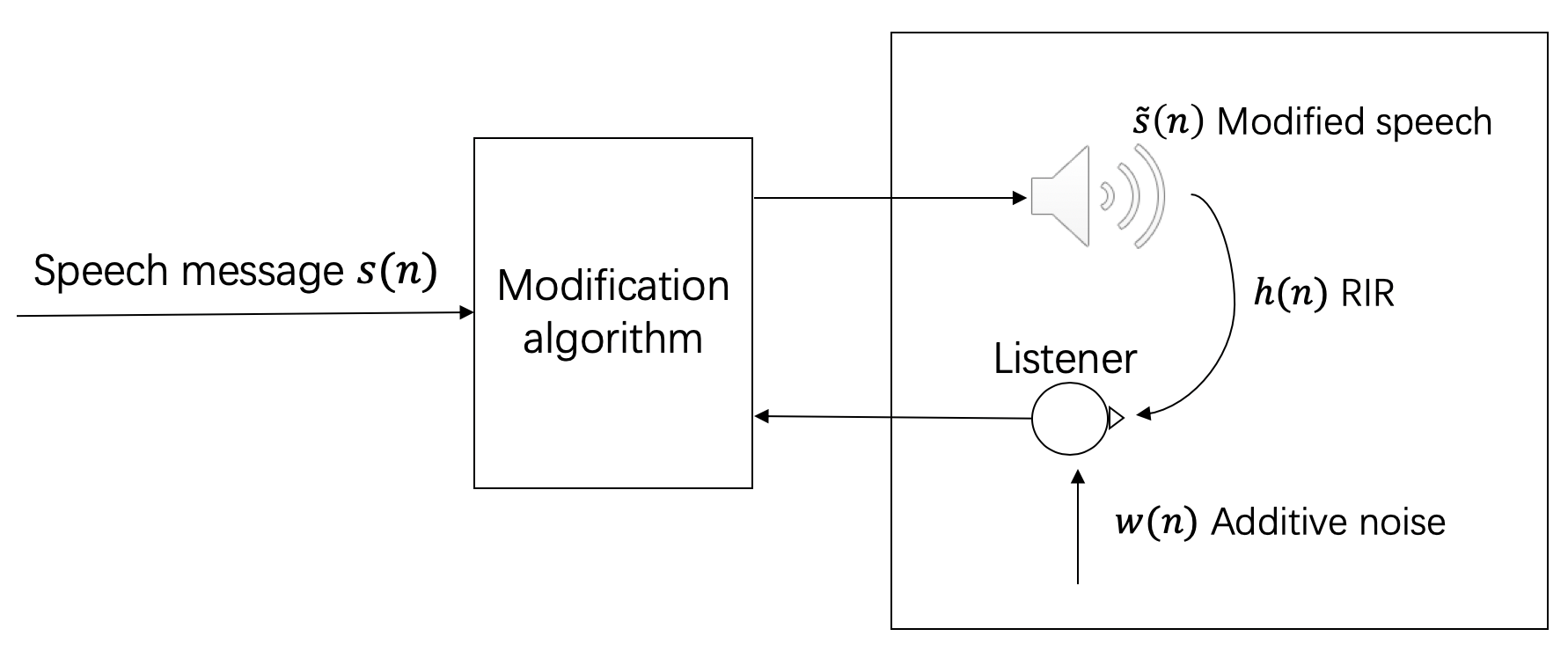}
  \caption{Schematic diagram of NELE scenario.}
  \label{fig:scenario}
    \vspace{-5mm}
\end{figure}

\section{Problem Formulation and Assumptions}
\label{sec:problemform}

\begin{figure*}[tbp]
  \centering
  \subfigure[\textit{D} training process]{
    \includegraphics[height=94.3pt,width=430pt]{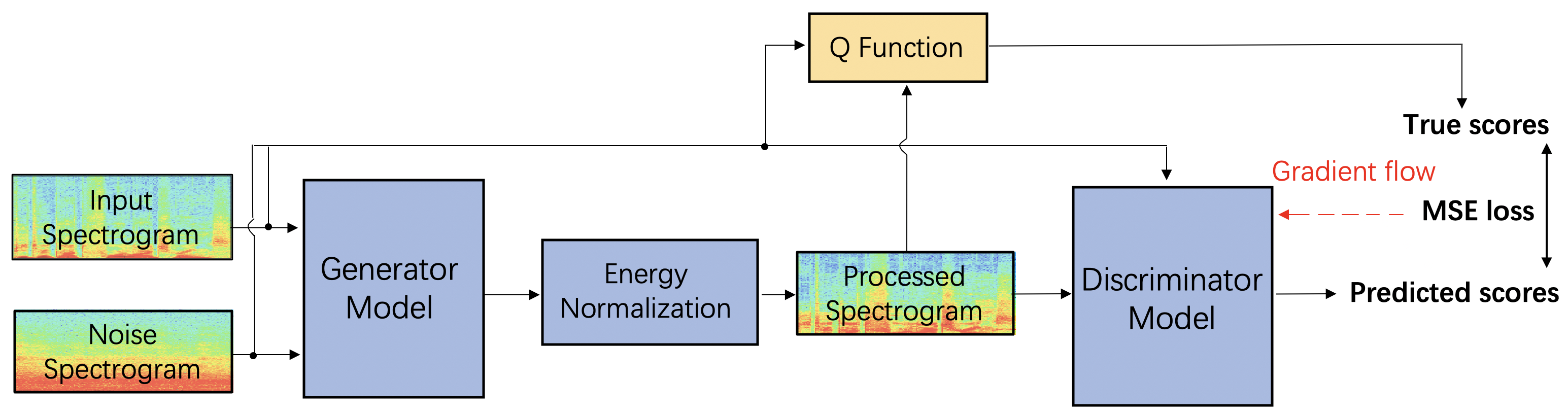}
  }
  \vspace{-2mm}
  \subfigure[\textit{G} training process]{
    \includegraphics[height=74.3pt,width=430pt]{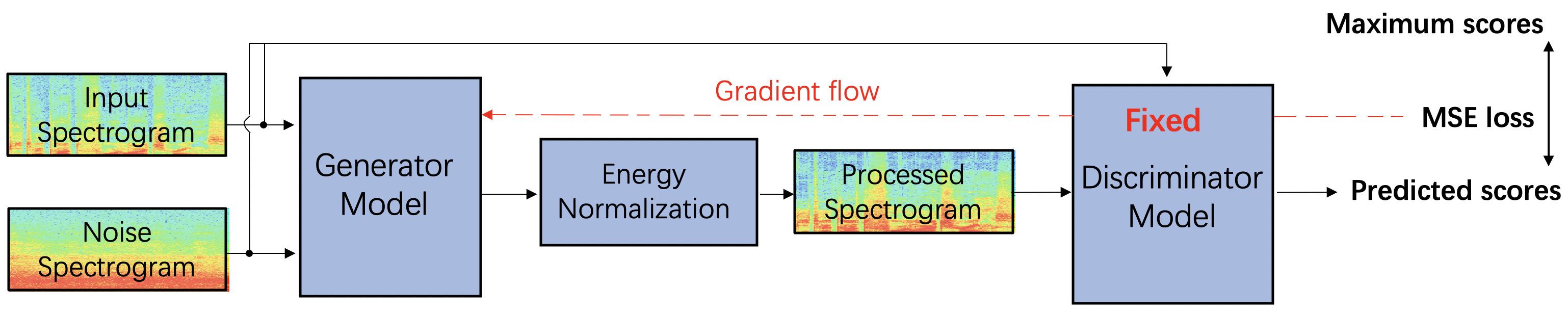}
  }
  \vspace{-2mm}
  \caption{The framework of iMetricGAN and its training process.}
  \label{fig:trainprocess}
  \vspace{-5mm}
\end{figure*}

Consider a real scenario of the NELE task, as depicted in Fig.~\ref{fig:scenario}.
Let $s(n)$ be the original speech signal. A modification algorithm is applied to $s(n)$ before it is played out by the loudspeaker and the processed output is denoted as $\tilde{s}(n)$. The observed signal $y(n)$ at the listener end is thus given by
\begin{equation}
  y(n) = h(n)*\tilde{s}(n) + w(n)
  \label{observedSig1}
\end{equation}
where $*$ denotes a convolution operation, $h(n)$ is the room impulse response (RIR)\footnote[1]{Loudspeaker response is integrated into RIR for simplicity.}, and $w(n)$ is the additive background noise. An assumption is made that the RIR $h(n)$ and the noise $w(n)$ can be estimated with an acoustic echo cancellation (AEC) technique \cite{gilloire1992adaptive}. Therefore, the general NELE task is formulated as finding an algorithm to modify the natural speech to improve its intelligibility in known noise and room conditions.

In this work, for simplicity, we take only the noise signal $w(n)$ into account and disregard $h(n)$, the influence of reverberation. Meanwhile, we do not change the RMS level and the duration of speech before and after modifications. With these assumptions and constraints, the problem can thus be formulated to design a DNN-based mapping function $\Phi(.)$, as
\vspace{-1.5mm}
\begin{equation}
    \begin{aligned}
    \tilde{s} = \Phi(s,w) \quad
    \mathrm{s.t.} \quad & RMS(\tilde{s}) = RMS(s),\\
    & Dur(\tilde{s}) = Dur(s)
    \label{dnnformulation}
    \end{aligned}
    \vspace{-1mm}
\end{equation}
where sample index $n$ is omitted from this point forward and \textit{Dur(.)} denotes the duration of the signal. It takes as input the unprocessed speech $s$ and the noise signal $w$. The output $\tilde{s}$ after modification is more intelligible when masked with noise.

\section{Proposed iMetricGAN Model}
In the field of speech enhancement, MetricGAN \cite{fu2019metricgan} has shown a powerful ability to optimize complex and even non-differentiable speech quality metrics, such as PESQ \cite{pesq}. The proposed \mbox{iMetricGAN} adapts and revises the original MetricGAN for the intelligibility enhancement task, where the target metric to be optimized is the speech intelligibility measure.

\subsection{Selecting intelligibility measures}
\label{sec:imeasure}
Objective measures are designed to predict the intelligibility score of speech. We select SIIB \cite{siib} and ESTOI \cite{estoi} measures as our optimization targets because they have achieved state-of-the-art performance (i.e., high correlations with listening tests), as demonstrated in \cite{van2018evaluation}. Both of these measures require reference and degraded signals as inputs for assessing relative intelligibility difference. To be specific, the reference is the original speech $s$, and the degraded signal is $\tilde{s}+w$. Objective intelligibility scores can thus be rated by the measures. Nevertheless, they cannot be directly set as the training targets for a DNN system, since both measures are quite complex and non-differentiable. Hence, we utilize \mbox{iMetricGAN} to overcome this obstruction.

\subsection{Model description and training process}
\label{sec:modeldescrip}

The model framework is depicted in Fig.~\ref{fig:trainprocess}. It consists of a generator (\textit{G}) network and a discriminator (\textit{D}) network. \textit{G} receives speech $s$ and noise $w$ and then generates the enhanced speech. An energy normalization layer is inserted to guarantee the energy is maintained after modification. The final processed speech is notated as $G(s,w)$. The cascading \textit{D} is utilized to predict the intelligibility score of the enhanced speech $G(s,w)$, given $s$ and $w$. The output of \textit{D} is notated as $D\left( G(s,w),s,w \right)$ and expected to be close to the true intelligibility score calculated by a specific measure. We introduce the function $Q(.)$ to represent the intelligibility measures to be modeled, i.e., SIIB and ESTOI. With the above notations, the training target of \textit{D}, shown in Fig.~\ref{fig:trainprocess} (a), can be represented to minimize the following loss function:
\vspace{-0.5mm}
\begin{equation}
    L_D = \mathbb{E}_{s,w} [( D(G(s,w),s,w) - Q(G(s,w),s,w))^2]
    \label{eq:ld1}
\vspace{-0.5mm}
\end{equation}
Moreover, we introduce $\hat{s}$, the signal example that is enhanced by reference modification algorithms such as SSDRC \cite{drc1}, in the \textit{D} training process. The loss function is thus extended to Equation~(\ref{eq:ld2}).
\vspace{-0.85mm}
\begin{equation}
\begin{aligned}
    L_D = \mathbb{E}_{s,w} [ & (D (G(s,w),s,w) - Q(G(s,w),s,w))^2\\
    & + (D(\hat{s},s,w) - Q(\hat{s},s,w))^2 ]
    \label{eq:ld2}
\end{aligned}
\vspace{-0.5mm}
\end{equation}
The motivation for introducing $\hat{s}$ is to improve the generalization of \textit{D}. By feeding it with not only $G(s,w)$ but also the signals modified by various other algorithms, \textit{D} is encouraged to predict the intelligibility scores in a more accurate way. Thus Equation~(\ref{eq:ld2}) can be seen as the loss function with auxiliary knowledge, while Equation~(\ref{eq:ld1}) is the loss function with zero knowledge. Note that $\hat{s}$ should not be regarded as the ground truth or the training label. In fact, the experimental results in Section~\ref{sec:result} demonstrate that \mbox{iMetricGAN} still works well even without introducing $\hat{s}$.

For the \textit{G} training process shown in Fig.~\ref{fig:trainprocess} (b), \textit{D}'s parameters are fixed and \textit{G} is trained to reach intelligibility scores as high as possible. To achieve this, the target score $t$ in Equation~(\ref{eq:lg}) is assigned to the maximum value of the intelligibility measure.
\vspace{-0.5mm}
\begin{equation}
    L_G = \mathbb{E}_{s,w}[(D(G(s,w),s,w) - t)^2]
    \label{eq:lg}
\end{equation}

\textit{G} and \textit{D} are iteratively trained until convergence. \textit{G} acts as an enhancement module and is trained to cheat \textit{D} in order to achieve a higher intelligibility score. On the other hand, \textit{D} tries to not be cheated and to accurately evaluate the score of the modified speech. This minimax game finally makes both \textit{G} and \textit{D} effective. Consequently, the input speech can be enhanced to a more intelligible level by \textit{G}. 

\section{Experimental Setup}
\label{sec:exp}

\subsection{Data preparation}
\label{sec:dataprep}
The data used in our experiments were provided by the Hurricane Challenge 2 (https://hurricane-challenge.inf.ed.ac.uk/). Three languages are considered (English, German, Spanish) and 290 matrix sentences are available (100 each for German and Spanish; 90 for English). Sentences were uttered by native speakers under 3 reverberation conditions (Near, Mid, and Far). For each reverberation, sentences were presented with 3 different SNRs corresponding to three intelligibility levels: approximately 25\%, 50\%, and 75\% correctly understood words. The detailed configuration is provided in Table~\ref{tab:rec}. Since we ignore the influence of reverberation, there are actually 9 SNRs for each sentence per language. All signals are sampled at 44.1 kHz and the masker signal is Cafeteria noise.
\begin{table}[t]
    \caption{SNRs (dB) of matrix sentences under different conditions}
    \label{tab:rec}
    \vspace{-2mm}
    \centering
    \begin{tabular}{p{0.06\textwidth}@{}l r r r r}
        \toprule
         \multicolumn{2}{c}{Intelligibility \quad} & 25\% & 50\% & 75\% \\
        \midrule 
          \multirow{3}{*}{English} &
           \quad Near & --12 & --7.5 & --3 \\
          &  \quad Mid & --4 & +2 & +8 \\
          &  \quad Far & --2 & +4 & +10 \\
          \midrule
          \multirow{3}{*}{German} &
           \quad Near & --15 & --12.5 & --10 \\
          &  \quad Mid & --9 & --6 & --3 \\
          &  \quad Far & --9 & --5 & --1 \\
          \midrule
        \multirow{3}{*}{Spanish} &
           \quad Near & --18.5 & --15.5 & --12.5 \\
          &  \quad Mid & --12 & --9 & --6 \\
          &  \quad Far & --12 & --8 & --4 \\
        \bottomrule 
    \end{tabular}
     \vspace{-5mm}
\end{table}

We chose German and Spanish speech as the training set, with a total of 900 sentences (100 sentences $\times$ 9 SNRs) for each language. Also, for data augmentation, we extracted 1,720 external sentences: 720 Spanish from the Sharvard corpus \cite{Sharvard} and 1000 German from the EMIME corpus \cite{EMIME}. Each of them was resampled to 44.1 kHZ and mixed with masker signals at six different SNR levels (randomly selected from 9 corresponding SNRs) in order to form 10,320 extra sentences. In total, 12,120 German and Spanish sentences were used for training. We did not include English speech as training data since we wanted to investigate a language-mismatched condition as well as language-matched conditions. The 810 English sentences (90 sentences $\times$ 9 SNRs) were used for testing only. To prepare the enhanced signal example $\hat{s}$ introduced in Section~\ref{sec:modeldescrip}, we selected three reference algorithms: (1) OptSII \cite{optSII}, a linear filter to maximize the Speech Intelligibility Index (SII), (2) OptMI \cite{optMI}, a linear filter to optimally redistribute energy based on mutual information criterion, and (3) SSDRC \cite{drc1}, a method to integrate spectral shaping and dynamic range compression. Each training sentence was randomly processed by one of these three algorithms to obtain its enhanced example. 

\subsection{Model architecture}

All input signals are first transformed to magnitude spectrograms by short-time Fourier transform (STFT). A 1024-point Hanning window with 512-point hop size is applied and results in 513 frequency bins. Power-law compression \cite{powerlaw} with parameter $p=0.3$ is followed to compress spectrograms. We concatenate the spectrograms of input speech and masker noise to form 1026-dimensional (513$\times$2) input features for \mbox{iMetricGAN}. After modification, the processed spectrogram is converted into a time-domain waveform using inverse STFT with the phase of the input speech.

The generator model \textit{G} is composed of two BLSTM layers, each with 400 hidden nodes, and two fully connected layers, each with 600 nodes. The activation function for the first fully connected layer is LeakyReLU with slope $=0.3$, and the last output layer is set as follows:
\begin{equation}
    output = \exp{(1.5 + 4 * tanh(m))} 
    \label{eq:output}
\end{equation}
where $m$ is the result of the previous layer. This output serves as scale factors, which are point-wise multiplied with the input spectrogram (unmodified speech) to produce an enhanced spectrogram. Scale factors modify the input speech by redistributing its energy: the T-F bin is boosted or declined with the corresponding scale value. The scale range of Equation~(\ref{eq:output}), which is approximately $0.08$ to $255$, is empirically chosen. We expect such a wide range will facilitate the processing ability of \textit{G}. Once processed by \textit{G}, the enhanced spectrogram is normalized by the \textit{energy normalization} layer, where the total squared energy of the output spectrogram is normalized to be the same as that of the input. The final processed spectrogram is sequentially passed on to the network \textit{D}.

As shown in Fig.~\ref{fig:trainprocess}, the input features for \textit{D} are 3-channel spectrograms, i.e., (processed, unprocessed, noise). \textit{D} consists of five layers of 2-D CNN with the following number of filters and kernel size: [8 (5, 5)], [16 (7, 7)], [32 (10, 10], [48 (15, 15)], and [64 (20, 20)], each with LeakyReLU activation. Global average pooling is followed by the last CNN layer to produce a fixed 64-dimensional feature. Two fully connected layers are successively added, each with 64 and 10 nodes with LeakyReLU. The last layer of \textit{D} is also fully connected and its output represents the scores of the intelligibility metrics. Therefore, the number of nodes in the last layer is equal to that of the intelligibility metrics we consider. For example, if we have \textit{D} predict SIIB and \mbox{ESTOI} scores simultaneously, it should be set to $2$. We normalize the SIIB score so that it ranges from 0 to 1, which is consistent with the range of the \mbox{ESTOI} score. Since both metrics of interest are bounded in \mbox{[0, 1]}, the sigmoid activation function is used in the last layer. Similar to \cite{fu2019metricgan}, all the layers in \textit{D} are constrained to be 1-Lipschitz continuous by spectral normalization \cite{miyato2018spectral} to stabilize the training process\footnote[2]{Source codes of this work are available at \texttt{https://github.\\com/nii-yamagishilab/intelligibility-MetricGAN}}.

\section{Results}
\label{sec:result}

\subsection{Notations of different modification methods}
As described in Section~\ref{sec:imeasure}, we have different options for learning metrics (SIIB, \mbox{ESTOI}, or both). In addition, two loss functions, Equations~(\ref{eq:ld1}) and (\ref{eq:ld2}), can be chosen in the training process, depending on the use of enhanced examples. Therefore, we built and compared the iMetricGAN model with three different variations\footnote[3]{Audio samples of the tested systems are available at \texttt{https://\\nii-yamagishilab.github.io/samples-iMetricGAN}}. To explain them, we use the following notations.
\begin{itemize}
    \item \textbf{SiibGAN-zs}: Learning target is SIIB, with Equation~(\ref{eq:ld1}) as the loss function. Since there is no enhanced example provided in this loss function, the model is trained in a \mbox{\textit{zero-short}} (zs) manner.
    \item \textbf{SiibGAN}: Learning target is SIIB, with Equation~(\ref{eq:ld2}) as the loss function.
    \item \textbf{MultiGAN}: Learning target includes multiple metrics, SIIB and ESTOI, with Equation~(\ref{eq:ld2}) as the loss function.
\end{itemize}

\subsection{Objective evaluations}

In addition to different iMetricGAN variants, three reference algorithms (OptMI, OptSII, SSDRC) used to produce enhanced examples were evaluated. Experimental results (SIIB and ESTOI) are presented in Table~\ref{tab:objresult}. 
\begin{table}[t]
    \caption{Average intelligibility scores on English test set under Cafeteria noise.}
    \label{tab:objresult}
    \vspace{-2mm}
    \centering
    \begin{tabular}{P{0.13\textwidth} P{0.07\textwidth} P{0.07\textwidth}}
        \toprule
         Algorithms & ESTOI & SIIB \\
        \midrule 
          Plain & 0.305 & 0.390 \\ 
          \midrule 
          OptMI & 0.423 & 0.591 \\
          OptSII & 0.416 & 0.623 \\
          SSDRC & 0.473 & 0.647 \\
          \midrule
          SiibGAN-zs & 0.386 & 0.660\\
          SiibGAN & 0.413 & \textbf{0.692}\\
          MultiGAN & \textbf{0.476} & 0.689\\
        \bottomrule 
    \end{tabular}
      \vspace{-6mm}
\end{table}

As shown, with modifications, the intelligibility scores of unmodified plain speech were effectively improved. Among all testing algorithms, MultiGAN had the best overall performance considering both \mbox{ESTOI} and SIIB scores, as expected. It surpassed the state-of-the-art SSDRC approach as well as OptMI and OptSII. By optimizing multiple metrics, it significantly outperformed another two \mbox{iMetricGAN} variants in terms of ESTOI, with only a slightly lower SIIB score compared to SiibGAN. For the \mbox{SiibGAN-zs} approach, the performance was degraded because the \textit{D} network cannot be well trained using a zero-shot approach. Even so, it brought a significant intelligibility gain to plain speech, which further demonstrates the effectiveness of our proposed \mbox{iMetricGAN} model.

\subsection{Subjective evaluations}

% Results may be further updated if outlier sample is detected by organizers.
Formal listening tests were conducted using the framework of the Hurricane Challenge 2. We submitted the entry of the MultiGAN algorithm, since it achieved the best performance in the objective evaluations. These listening tests took into account the sentences of three languages (English, German, Spanish) combined with different SNR and reverberation conditions. The SNRs used in the final tests were slightly different from those listed in Table~\ref{tab:rec} due to technical reasons\footnote[4]{Hurricane Challenge organizers adjusted the SNRs to allow for sufficient headroom in which the entries could show their performance.}. Note that all provided languages and reverberations were considered in the subjective tests, while only English sentences without reverberation were used for the objective evaluation described in the previous section. We returned only the modified (enhanced) speech to the challenge organizers. These modified speech signals were remixed with noise and RIR by organizers, and then evaluated by native listeners. The masker signal was still Cafeteria noise but not a sample-by-sample equivalent of the signals we used in the training phase. Specifically, more than 180 listeners ($>$60 for each language) were asked to type in what they heard in the experiments. Word accuracy rate, i.e., the percentage of correct words in a transcription, was then calculated as the performance measure of intelligibility. 

\begin{figure}[t]
  \centering
  \subfigure[Word accuracy in English]{
    \includegraphics[width=\linewidth]{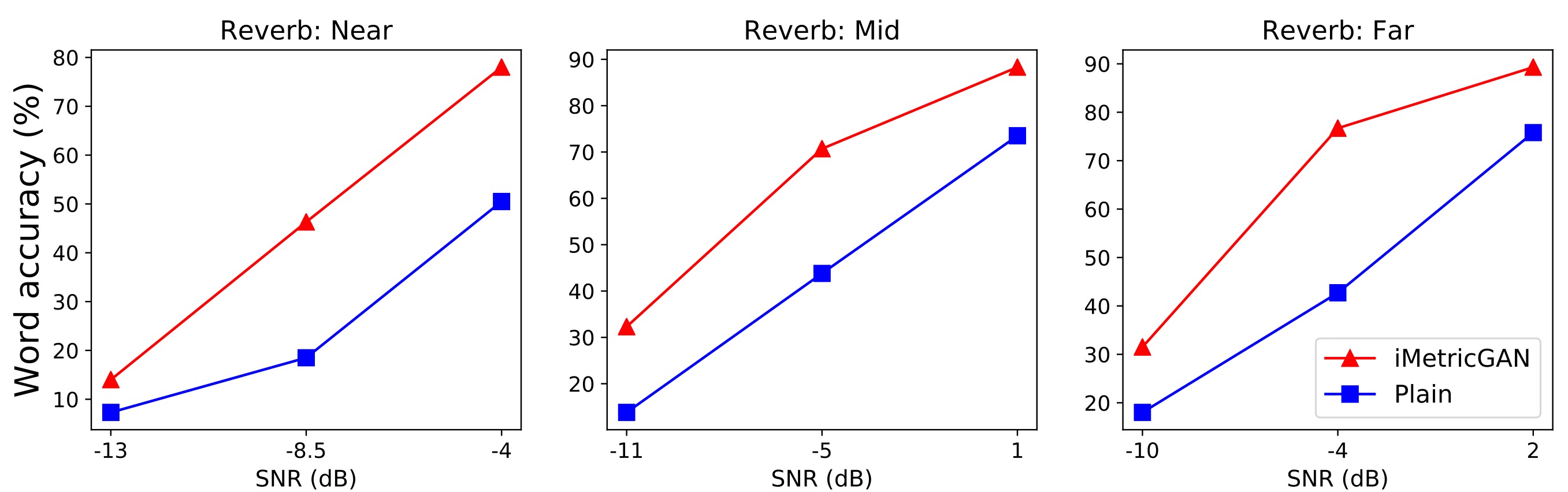}
  }
  \subfigure[Word accuracy in German]{
    \includegraphics[width=\linewidth]{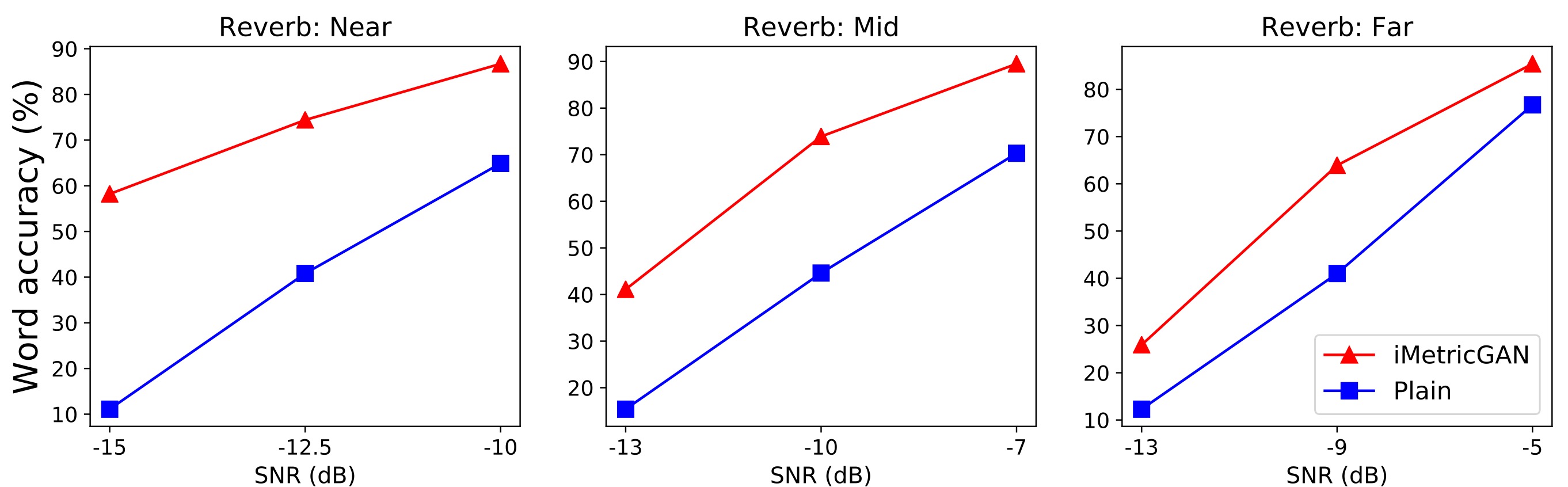}
  }
  \subfigure[Word accuracy in Spanish]{
    \includegraphics[width=\linewidth]{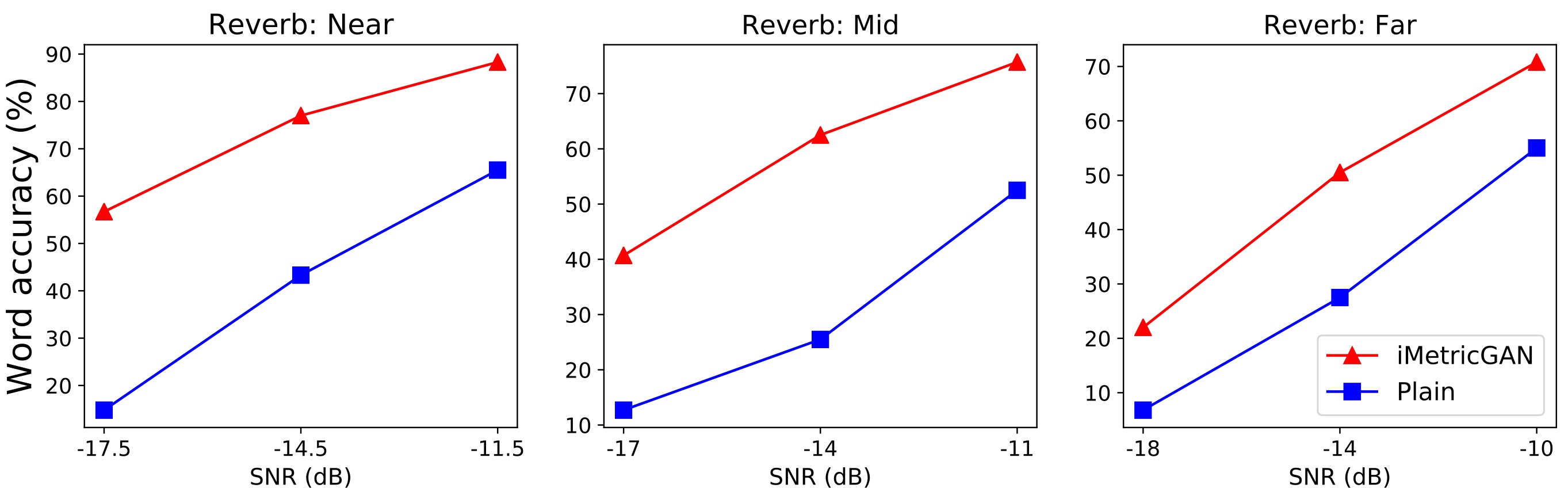}
  }
    \vspace{-4mm}
  \caption{Average word accuracy for plain and iMetricGAN-based modified speech in three languages under different SNR and reverberation conditions.}
  \label{fig:sbjres}
  \vspace{-7mm}
\end{figure}

Figure~\ref{fig:sbjres} shows the results for different listening conditions. Our proposed \mbox{iMetricGAN} achieved significant intelligibility gains in each language under all SNR and reverberation conditions. Note that we disregarded the influence of RIR when building the \mbox{iMetricGAN} model, but it still worked quite well in reverberant environments.

\subsection{Discussions}
\vspace{-1mm}
In Fig.~\ref{fig:sbjres} (a), we can see that the word accuracy was improved for the English test set even though English sentences were not included in the training set. This demonstrates the good generalization capability of the \mbox{iMetricGAN} model, which can generalize well to mismatched language, speaker, and SNR level conditions. Future work will include the real-time implementation of \mbox{iMetricGAN}. To achieve this, we should change the original BLSTM to the uni-directional version and use noise power spectral density (PSD), which can be feasibly estimated as input noise information, instead of a raw noise signal. In the real-time inference stage, no global energy constraint can be guaranteed since the future signal values are not available, while the volume of speech can be adaptively maintained in a proper way by utilizing the automatic gain control (AGC) technique \cite{shan1988adaptive}. Another future direction will involve introducing more advanced intelligibility metrics such as HASPI \cite{kates2014hearing} and HEGP \cite{tang2016glimpse} to the model training. We also plan to investigate ways of integrating speech quality metrics such as PESQ to enhance the quality of the modified speech.

\vspace{-2mm}
\section{Conclusion}
\vspace{-0.3mm}
In this paper, we proposed the iMetricGAN model to enhance the intelligibility of speech-in-noise. Objective results show that our approach outperforms the state-of-the-art SSDRC method in terms of SIIB and ESTOI scores. Large-scale formal listening tests further show its effectiveness in intelligibility enhancement across different languages and background environment conditions.

\vspace{0.2mm}

\noindent
\textbf{Acknowledgments}
%\footnotesize{
This work was partially supported by a JST CREST Grant (JPMJCR18A6, VoicePersonae project), Japan, and by MEXT KAKENHI Grants (16H06302, 17H04687, 18H04120, 18H04112, 18KT0051, 19K24372), Japan. The numerical calculations were carried out on the TSUBAME 3.0 supercomputer at the Tokyo Institute of Technology.
%}

\bibliography{template.bbl}

\end{document}